\documentclass[12pt,english,a4paper]{article}

\usepackage{amssymb,amsfonts,amsthm,amsmath,graphicx,xspace,colortbl,bm,lineno}
\usepackage[longnamesfirst]{natbib}
\usepackage[figuresright]{rotating}
\usepackage{tabularx}
\usepackage{times}
\usepackage{bbold}
\usepackage{babel}
\usepackage[printfigures]{figcaps}
\usepackage[font=small,labelfont=bf,tableposition=top]{caption}
\usepackage[colorlinks=true,linkcolor=blue,citecolor=blue,urlcolor=blue,pdftitle={On site occupancy models with heterogeneity},pdfauthor={Wen-Han Hwang, Jakub Stoklosa, Lu-Fang Chen},pdfsubject={manuscript},pdfkeywords={NB}]{hyperref}

\usepackage{xr}
\usepackage{changes}
\usepackage{footnote}
\makesavenoteenv{tabular}
\makesavenoteenv{table}
\usepackage{placeins} %Allows barriers to be placed for tighter control of figure position
\usepackage{authblk} %Allows author to have block and multiple affiliations
\usepackage{threeparttable}

\newcommand*\patchAmsMathEnvironmentForLineno[1]{%
\expandafter\let\csname old#1\expandafter\endcsname\csname #1\endcsname
\expandafter\let\csname oldend#1\expandafter\endcsname\csname end#1\endcsname
\renewenvironment{#1}%
{\linenomath\csname old#1\endcsname}%
{\csname oldend#1\endcsname\endlinenomath}}% 
\newcommand*\patchBothAmsMathEnvironmentsForLineno[1]{%
\patchAmsMathEnvironmentForLineno{#1}%
\patchAmsMathEnvironmentForLineno{#1*}}%
\AtBeginDocument{%
\patchBothAmsMathEnvironmentsForLineno{equation}%
\patchBothAmsMathEnvironmentsForLineno{align}%
\patchBothAmsMathEnvironmentsForLineno{flalign}%
\patchBothAmsMathEnvironmentsForLineno{alignat}%
\patchBothAmsMathEnvironmentsForLineno{gather}%
\patchBothAmsMathEnvironmentsForLineno{multline}%
}

\def\bvec#1{\mbox{\boldmath $#1$}}

\def\btheta{{\bvec \theta}}
\def\bgamma{{\bvec \gamma}}

\def\bb0{{\bvec 0}}
\def\rI{\mathrm{\bf I}}
\def\tb#1#2{\mathop{#1\vphantom{\sum}}\limits_{\displaystyle #2}}

\def\rI{{\rm I}}
\newcommand{\ie}{\emph{i.e.,}\ }
\newcommand{\eg}{\emph{e.g.,}\ }

\newtheorem{theorem}{Theorem}%[section]

\newtheorem{proposition}[theorem]{Proposition}

\definecolor{ao(english)}{rgb}{0.0, 0.5, 0.0}

\title{On site occupancy models with heterogeneity}

\author{Wen-Han Hwang$^1$\footnote{Corresponding author Email: wenhan@nchu.edu.tw}, Jakub Stoklosa$^2$ and Lu-Fang Chen$^3$ \\
\small $^1$ Institute of Statistics, National Chung Hsing University, Taiwan\\
\small $^2$ School of Mathematics and Statistics and Evolution \& Ecology Research Centre, \\
\small The University of New South Wales, Australia.\\
\small $^3$ Teaching Center of Natural Science, Minghsin University of Science and Technology, Taiwan\\
}

\begin{document}

\maketitle

\begin{abstract}
Site occupancy models are routinely used to estimate the probability of species presence from either abundance or presence--absence data collected across sites with repeated sampling occasions. In the last two decades, a broad class of occupancy models have been developed, but little attention has been given in examining the effects of heterogeneity in parameter estimation. This study focuses on occupancy models where heterogeneity is present in detection intensity and the presence probability. We show that the presence probability will be underestimated if detection heterogeneity is ignored. On the other hand, the behaviour is different if heterogeneity in the presence probability is ignored; notably, an estimate of the average presence probability may be unbiased or over- or under-estimated depending on the relationship between detection and presence probabilities. In addition, when heterogeneity in the detection intensity is related to covariates, we propose a conditional likelihood approach to estimate the detection intensity parameters. This alternative method shares an optimal estimating function property and it ensures robustness against model specification on the presence probability. We then propose a consistent estimator for the average presence probability, provided that the detection intensity component model is correctly specified. We illustrate the bias effects and estimator performance in simulation studies and real data analysis.
\end{abstract}

\noindent{\bf Key words}: Conditional likelihood; Detection and presence probability heterogeneity; Mixture models; Zero-inflated Poisson distribution.

\section{Introduction}
\label{sec:intro}

Acquiring accurate estimates of the probability of species presence (a.k.a.\ the occurrence/occupancy rate) is of fundamental importance to ecologists and environmental conservationists, as they provide a useful measure of the species' population status and conservation management \citep{wenger}. A popular strategy is to use occupancy modelling, which focuses on inference about species distribution over space where temporal or spatial replication allows for imperfect detection of objects in the survey. The usual sampling protocol consists of multiple visits to sites, and the species of interest are noted as being present or not present at each site. Additional information, such as environmental covariate data, can also be collected at each site and used to model heterogeneity. In site occupancy modelling, heterogeneity arises from: variation in abundance and site characteristics; heterogeneous sampling effort in space; and other unaccounted factors \citep{royle, louvriera}.

The observed response data from such sampling usually consists of abundance/count data (\ie species' frequency) or presence--absences (\ie a binary outcome). To estimate the presence probability, the use of the {\it site occupancy model} \citep{macKenzie} is by far the most popular. Generally, a detection component is included in these models to account for the incomplete detection arising from sampling. Thus, site occupancy models will primarily constitute of: (1) an occurrence component which is also referred to as the presence/occurrence/occupancy rate or probability, and (2) a detection component which is also referred to as a detection probability or intensity. Throughout this paper, we follow \citet{macKenzie} and use the terms presence probability and detection intensity or probability. In the last two decades, a broad class of occupancy models have now been developed \citep[for a recent review, see][]{denes, guillera(2016), macKenzie(2017)}. 

In this study, we investigate several fundamental properties for site occupancy models when detection or presence probabilities are subject to heterogeneity \citep{turek}. For simplicity, we uses the term heterogeneity when detection or presence probabilities are not constant. We consider site occupancy abundance models based on a Poisson modelling framework \citep{wenger} which turns out to be a zero-inflated Poisson model \citep{welsh, martin}. Under this setting, we begin by showing that a conditional maximum likelihood estimator is identical to the conventional maximum likelihood estimator when the detection intensity follows a mixing distribution. \citet{sanathanan} showed that these two approaches are asymptotically equivalent when estimating the size from truncated samples, and \citet{fewster(2009)} and \citet{schofield(2016)} similarly showed this for mark--recapture/binomial models. Interestingly, an analogous conclusion when detection intensities are related to covariates has yet to be established in the literature. We prove that these two approaches are not asymptotically equivalent in this circumstance. Moreover, we find that conditional likelihood benefits from an optimal estimating function property and ensures robustness against model specification on the presence probability.

We address the effects on presence probability estimation when heterogeneity in detection and presence probabilities are ignored. We show that if detection heterogeneity is unaccounted for, the presence probability will be underestimated, and an approximate bias is also derived. Secondly, when presence probabilities are related to covariates, we find that the impacts on presence probability estimation can be vastly different if heterogeneity in presence probabilities is ignored. Specifically, the average presence probability estimate may be unbiased, or over- or under-estimated depending on the relationship between detection and occurrence. 

We also propose a robust method based on the conditional likelihood where the average presence probability can still be consistently estimated so long as the detection component model is correctly specified. In summary, we found that model specification is crucial for the detection intensity/probability component but less critical for the presence probability component. Our results are then extended to occupancy models when using multiple visit presence--absence data where we base our modelling framework on a zero-inflated binomial model~\citep{macKenzie}.

The underestimation of presence probabilities due to a lack of detection heterogeneity has been observed in \citet{royle03} and \citet{mcclintock} who identified this phenomenon in presence--absence occupancy models. However, these effects were only demonstrated through simulation studies or by parodying the concept of heterogeneity among individuals, as observed in capture--recapture models \citep{royle03}. Currently, there is no theoretical justification for this observation. Our second findings which examined the impacts of ignoring heterogeneity in presence probabilities turns out to be an interesting phenomena, and has yet to be considered in the literature.

In Section~\ref{sec:model} we describe site occupancy models and parameter estimation based on a Poisson framework, and show that the conditional likelihood estimator is asymptotically equivalent to the maximum likelihood estimator when the presence probability is a constant. We then investigate the effects of ignoring heterogeneity in detection or presence probabilities when using regression models. Section~\ref{sec:binary} extends the framework to presence--absence type data based on a zero-inflated binomial model. Simulations and real-data examples are given in Sections~\ref{sec:sim} and~\ref{sec:ex}, respectively. A discussion of results and some future work are given in Section~\ref{sec:dis}. Proofs of all propositions are given in a Web Appendix.

\section{Abundance data: Poisson models}
\label{sec:model}
 
We assume that there are $n$ sites where each site is visited $T$ times. Let $y_{ij}$ be the number of observed/detected individuals of a species at site $i = 1, \ldots, n$ {and visit $j = 1, \ldots, T$. Let $y_{i} = \sum_{j = 1}^T y_{ij}$ be the total frequency of a detected individuals at site $i$.} Let $A$ be the collection of distinct sites with a non-zero observed total frequency -- \ie $A = \{i: y_{i} \geq 1\}$, and let $A^c$ be the complement of event $A$. Let $\psi$ be the probability that each site is occupied/present by the species. We call $\psi$ the presence probability, also commonly referred to as the occurrence probability and occupancy parameter. In general, when fitting site occupancy models, the presence probability is of key interest to ecologists~\citep{macKenzie}.

If site $i$ is not occupied by the species, then $y_{ij} = 0$ for all $j = 1, \ldots, T$ (as does $y_i = 0$), definitely. However, if the species occupied site $i$, then $y_{ij}$ is a non-negative count integer. Therefore, a popular framework for such count data assumes that $y_{ij}$ is Poisson distributed \citep{welsh, royle03}. Again, under the common independence assumption between 
visits, the distribution of $y_i$ is also Poisson, denoted as $y_i \sim {\rm Poisson}(\lambda)$ where $\lambda$ represents the (homogeneous) detection intensity. Equivalently, under this framework, we can aggregate counts over visits since the model is still applicable if $T = 1$.

As we cannot observe the status of occupancy for each site, we can alternatively assume that $y_{i}$ follows a zero-inflated Poisson distribution~\citep{welsh, wenger} for all $i = 1, \ldots, n$, in this case we write $y_{i} \sim (1 - \psi) \rI(0) + \psi {\rm Poisson}(\lambda)$ where $\rI(\cdot)$ denotes the usual indicator function. The degree of detection intensity is reflected by the strength of combining abundance intensity and sampling effort at each site. Under certain conditions, it is possible to separate abundance intensity and sampling effort \citep{lele}, however we do not consider this in our study.

\subsection{Estimation}
\label{subsec:est}
Let $m_k$ be the number of sites that the species is observed in $k$ times for all $k \geq 0$, $m_k = \sum_{i = 1}^n\rI(y_i = k)$. Also, let $m_+ = \sum_{k \geq 1} m_k = n - m_0$ denote the number of sites with at least one observation which is equal to the size of $A$. The likelihood function under the zero-inflated Poisson distribution is given by
\begin{linenomath*}
\begin{eqnarray}
L(\lambda, \psi)
% & = & \prod_{i \in A} \left\{\psi\frac{\lambda^{y_{i}}e^{-\lambda}}{y_{i}!}\right\} \times \left\{(1 - \psi) + \psi e^{-\lambda}\right\}^{m_0} \nonumber \label{eq1}\\
& \propto & \prod_{k \geq 1} \left(e^{-\lambda}\lambda^k\right)^{m_k} \psi^{m_+} \times \left \{(1 - \psi) + \psi e^{-\lambda}\right\}^{m_0}.\label{eq2}
\end{eqnarray}
\end{linenomath*}

Throughout, we will refer to the above likelihood function (\ref{eq2}) as the {\it homogeneous abundance model}.

In practice, the detection intensity may vary across sampled sites due the variation in abundance and sampling effort. To allow for heterogeneous detection, we first consider a mixture distribution for $\lambda$ \citep{royle}. Specifically, let $g_\btheta(\lambda)$ be the probability density function (or probability mass function), and let $p_k(\btheta) = \int \frac{\lambda^ke^{-\lambda}}{k!} g_\btheta(\lambda) {\rm d}\lambda$ for all $k \geq 0$. Several well-known mixture distributions can be considered for $g_\btheta(\lambda)$ -- \eg finite mixtures~\citep{bohning} and gamma mixtures~\citep{white} where the latter yields a negative binomial distribution. Consequently, the likelihood function with a mixture distribution for the detection intensity is given by
\begin{linenomath*}
\begin{equation}
L(\btheta, \psi) = \left\{\prod_{k \geq 1} p_k(\btheta)^{m_k}\right\} \psi^{m_+} \times\{(1 - \psi) + \psi p_0(\btheta)\}^{m_0}. \label{eq4}
\end{equation}
\end{linenomath*}

The estimation procedure for $\btheta$ and $\psi$ can be established by maximizing the likelihood~(\ref{eq4}) -- \ie the maximum likelihood (maximum likelihood) estimate. Here, we consider an alternative estimating procedure that uses the conditional likelihood (conditional likelihood) as follows. This procedure has two stages:~(1) we first estimate $\btheta$ based on the conditional likelihood; and~(2) obtain an estimate of $\psi$ (using the estimate of $\btheta$). To see this, let $p_+(\btheta) = 1 - p_0(\btheta) = \sum_{k\geq 1} p_k(\btheta)$ and write the likelihood function~(\ref{eq4}) as
\begin{linenomath*}
\begin{eqnarray*}
L(\btheta, \psi) & \propto & L_1(\btheta) \times L_2(\btheta, \psi)\nonumber\\
& = & {m_+ \choose m_1, m_2, \ldots}\prod_{k \geq 1}\left\{\frac{p_k(\btheta)}{p_+(\btheta)}\right\}^{m_k} \times{n \choose m_+} \left\{\psi p_+(\btheta)\right\}^{m_+} \left\{(1 - \psi p_+(\btheta)
\right\}^{m_0},
\end{eqnarray*}
\end{linenomath*}
where $L_1(\btheta)$ is the conditional likelihood given $m_+$ sites are observed with at least one individual, and $L_2(\btheta, \psi)$ is the marginal likelihood of $m_0$. Since the conditional likelihood does not involve $\psi$, we can maximize $L_1(\btheta)$ with respect to $\btheta$ and obtain an estimate of $\btheta$. The resulting estimator is denoted by $\widehat\btheta$ which is then substituted into $L_2(\btheta, \psi)$ to obtain an estimator for the presence probability, given by $\widehat\psi = {m_+}/\{np_+(\widehat \btheta)\}$. %Hereafter, we denote $\widehat\psi$ as the conditional likelihood estimate of $\psi$.

%Moreover, under certain regularity conditions, both $\widehat\btheta$ and $\widehat \psi$ are consistent estimators. \comm{Wen-Han can you please justify this and add a reference that supports this claim.} To obtain standard errors, we use ${\rm Var}(\widehat \btheta)$ calculated from the inverse of observed Fisher information matrix of $L_1(\btheta)$. The standard error estimate of $\widehat \psi$ can be established by the delta method.

Lemma~1 (Web Appendix) provides a proof that the conditional likelihood approach yields the same estimate when using maximum likelihood. Hence, $\widehat\btheta$ and $\widehat\psi $ are also the maximum likelihood estimators. For multiple visit presence--absence data, this property has been shown by \citet{huggins20} based on the constant binomial model.

\subsection{Effects of ignoring detection heterogeneity}
\label{subsec:eff_det}

Next, we discuss the implications of ignoring heterogeneity when estimating the presence probability $\psi$. Let $\widehat\psi_0$ be the estimator of $\psi$ obtained via the homogeneous abundance model~(\ref{eq2}). %{The proof for the following proposition is given in the Web Appendix.}

%In the next proposition, we give the asymptotic bias of $\widehat \psi_0$ in the presence of detection heterogeneity.

\begin{proposition}
\label{prop3}
Under the general abundance model~(\ref{eq4}), if detection heterogeneity is ignored then the resulting estimator $\widehat\psi_0$ incurs an asymptotic downward bias only if $n \rightarrow \infty$. Moreover, the approximate bias is $\rho\psi$ where $\rho = \{1 + 0.5\sigma^2 /(e^{\mu} - 1 - \mu)\}^{-1}$ with $\mu$ and $\sigma^2$ are the respective mean and variance of $g_\btheta(\lambda)$.
\end{proposition}
The approximate bias of Proposition~1 reveals how the bias is related to the degree of detection heterogeneity ($\sigma^2$) and the mean intensity ($\mu$). For example, the bias would be negligible if $\mu$ is large; otherwise, the bias may be severe. We verify the bias expression in through a simulation study in Section~\ref{sec:sim}.
 
%\comm{The example can be considered to delete}\\
%{\bf Example:} Suppose that $\lambda$ follows an exponential distribution with mean $\mu$, then the probability function is $p_k(\mu) = \mu^k/(1 + \mu)^{k + 1}$ for all $k \ge 0$ -- \ie geometric distribution. Let $\widehat\lambda_0$ and $\widehat\psi_0$ denote the conditional maximum likelihood estimators of model (\ref{eq2}). It is easy to see that $\widehat\psi_{0} = {m_+}/\{n(1 - e^{-\widehat\lambda_0}) \}$, and $\widehat\lambda_0$ solves
%\begin{linenomath*}
%\begin{equation}
%\frac{\lambda}{1 - e^{-\lambda}} = \frac{\sum_{k \geq 1} km_k}{m_+}. \label{e31a}
%\end{equation}
%\end{linenomath*}

%Notice that the right hand side of (\ref{e31a}) converges in probability to $\mu/\{1 - p_0(\mu)\} = 1 + \mu $, so the limit of $\widehat\lambda_0$ solve for $\lambda/(1 - e^{-\lambda}) = 1 + \mu$. Clearly, the right hand side of (\ref{e31a}) is greater than $\mu/(1 - e^{-\mu})$ for $\mu > 0$. Hence, by the monotonicity of $x/(1 - e^{-x})$ for $x > 0$, we have $\widehat\lambda_0 \ge \mu$ (as does $\widehat\psi_0 \le \psi$) asymptotically, with probability one. 

\subsection{Regression models}
\label{subsec:reg}

Heterogeneity in detection intensity is often modelled with covariates in a regression setting -- \eg the detection intensity of site $i$ is modelled as $\lambda_i = \exp(\btheta'x_i)$ where $x_i$ is a vector of covariates and $\btheta$ is the associated regression parameter. Suppose that $y_i \sim (1 - \psi)\rI(0) + \psi {\rm Poisson}(\lambda_i)$, then the likelihood function is
\begin{linenomath*}
\begin{equation}
L(\btheta, \psi) = \prod_{i \in A}\left(\psi\frac{\lambda_i^{y_{i}} e^{-\lambda_i}}{y_{i}!}\right) \times \prod_{i \in A^c} \left\{(1 - \psi) + \psi e^{-\lambda_i}\right\}. \label{reglik} 
\end{equation}
\end{linenomath*}

Unlike the previously mentioned models, the conditional likelihood and maximum likelihood estimators under model (\ref{reglik}) are different. The conditional likelihood function is defined as 
\begin{linenomath*}
\begin{equation} 
L_1(\btheta) = \prod_{i \in A}\left(\frac{\lambda_i^{y_{i}} e^{-\lambda_i}}{1 - e^{-\lambda_i}}\right), \label{reglik2}
\end{equation}
\end{linenomath*}
which is independent of $\psi$. The conditional likelihood estimator $\widehat \btheta_c$ is the maximizer of (\ref{reglik2}). Denote $\widehat \lambda_i = \exp(\widehat\btheta_c' x_i)$, then we can estimate $\psi$ via the profile likelihood function $L_2(\psi) \propto \psi^{m_+} \prod_{i \in A^c} \left(1 - \psi + \psi e^{-\widehat\lambda_i}\right)$. Thus, we define the conditional likelihood estimator $\widehat \psi_c$ by solving 
\begin{linenomath*}
\begin{equation}
\frac{m_+}{\psi} - \sum_{i \in A^c} \frac{1 - e^{-\widehat \lambda_i}}{1 - \psi + \psi e^{-\widehat \lambda_i}} = 0. \label{psi-c}
\end{equation}
\end{linenomath*}

Let $(\widehat \btheta, \widehat\psi)$ be the maximum likelihood estimators of $(\btheta, \psi)$. Also, let ${\rm var}(\widehat\btheta)$ and ${\rm var}(\widehat\btheta_c)$ denote the corresponding asymptotic variance of $\widehat\btheta$ and $\widehat\btheta_c$, respectively. Similar notations are defined for ${\rm var}(\widehat\psi)$ and ${\rm var}(\widehat\psi_c)$. Let $\rI_i = \rI(y_i > 0)$, $\pi_i = 1 - e^{-\lambda_i}$, $g_i = y_i - \lambda_i/\pi_i \rI_i$, $h_i = (\rI_i - \psi \pi_i)/\{\psi(1 - \psi \pi_i)\}$ and $\phi_i = (g_i, h_i)'$. In Lemma~2 of the Web Appendix, we show that both estimating equations for $(\widehat \btheta, \widehat\psi)$ and $(\widehat \btheta_c, \widehat\psi _c)$ belong to the family of weighted estimating equations based on $\phi_i$. Our main results show the comparison between $(\widehat \btheta, \widehat\psi)$ and $(\widehat \btheta_c, \widehat\psi_c)$. %{The proof for the following proposition is given in the Web Appendix.}

\begin{proposition}
\label{prop5}
We have ${\rm var}(\widehat\btheta) \le {\rm var}(\widehat\btheta_c)$ in the sense that ${\rm var}(\widehat\btheta_c) - {\rm var}(\widehat\btheta)$ is a non-negative definite matrix. In addition, ${\rm var}(\widehat\psi) \le {\rm var}(\widehat\psi_c)$. Both of the equalities hold if $x_i$ is a constant for all $i$. Moreover, the score function for $(\widehat \btheta, \widehat\psi)$ is the optimal estimating function for $(\btheta, \psi)$. However, if $\psi$ is regarded as a nuisance parameter, then the conditional score function of $\widehat \btheta_c$ is the optimal estimating function for $\btheta$. 
\end{proposition}

The above proposition shows that the conditional likelihood estimator is less efficient than the usual maximum likelihood estimator (unless the covariate $x_i$ is a constant). The optimal property of the maximum likelihood approach is a fundamental result of \citet{godambe60}. On the other hand, the optimal property of the conditional score function for $\widehat\btheta_c$ is a consequence of \citet{godambe76}, which implies that $\widehat\btheta_c$ preserves some robustness in model misspecification for estimating $\btheta$. Moreover, the optimality holds if the presence probability is not a constant \ie the presence probability is $\psi_i$ for each $i$ and all $\psi_i$ are unknown parameters, see model (\ref{model:reg2}) below.

Also, there is a downward bias effect on estimating the presence probability if the regression model (\ref{reglik}) is misidentified as a homogeneous model (\ie no non-constant regressors are included). The proof is omitted as it is similar to the proof for Proposition~1, but $x_i$ is assumed to be collected from a random sample. A generalisation for the regression model is that the downward bias effect remains only if some of the covariates are ignored. We demonstrate this in a simulation study, see Section~\ref{sec:sim}.

%\begin{proposition}
%\label{prop6}
%
%Consider model~(\ref{reglik}) and assume that if detection heterogeneity is ignored and the estimate for $\psi$ is obtained via the homogeneous model~(\ref{eq2}), then the resulting estimator $\widehat\psi_0$ incurs an asymptotic downward bias only if $n$ is sufficiently large.
%\end{proposition}
%

\subsection{Effects of ignoring heterogeneity in the presence probability}
\label{subsec:eff_occ}

Finally, we investigated the effects of ignoring heterogeneity in presence probabilities. Since the status of presence is represented by a binary variable, heterogeneity in presence probabilities is generally modelled by a logistic regression model where the source of heterogeneity is represented by covariate information taken from sites. For convenience, denote $p_y (\btheta)$ as a mixture Poisson distribution with parameter $\btheta$, and assume that
\begin{linenomath*}
\begin{equation}
y_i \sim (1 - \psi_i) \rI(0) + \psi_i p_y (\btheta), \label{model:psi}
\end{equation} 
\end{linenomath*}
where $\psi_i = H(\gamma'z_i)$ for some covariates $z_i$ and $H(x) = 1/\{1 + \exp(-x)\}$ is the logistic function. Model~(\ref{model:psi}) is reduced to model~(\ref{eq4}) when $z_i = 1$ for all $i$.

Suppose that we are interested in the average presence probability $\bar \psi$ (\ie the average of $\psi_i$), we show that $\bar \psi$ can be consistently estimated even if heterogeneity in $\psi_i$ is ignored. %{The proof for the following proposition is given in the Web Appendix.}

\begin{proposition}
\label{prop7} 

Under model~(\ref{model:psi}), if heterogeneity in the presence probability is ignored and the estimate for $\psi$ is obtained via the model~(\ref{eq4}), then the resulting estimator converges to $\bar \psi$ in probability as $n$ increases to infinity.
\end{proposition} 

An extension for the above is not entirely correct when we consider a regression framework for the detection component. Specifically, we consider
\begin{linenomath*}
\begin{equation}
y_i \sim (1 - \psi_i) \rI(0) + \psi_i {\rm Poisson}(\lambda_i) \label{model:reg2}
\end{equation} 
\end{linenomath*}
where $\lambda_i = \exp(\btheta'x_i)$ and $\psi_i = 1/\{1 + \exp(-\gamma'z_i)\}$ for some covariates $x_i$ and $z_i$. 

Under this setting, we find that $\widehat\psi_c$ preserves consistency for $\bar \psi$ if detection intensities $\lambda_i$ and presence probabilities $\psi_i$ are uncorrelated (\eg the non-intercept covariates in $z_i$ are unrelated to those of $x_i$). However, it tends to overestimate $\bar \psi$ when $\lambda_i$ and $\psi_i$ are positively related, and underestimates $\bar \psi$, vice versa. The condition in Proposition~\ref{prop7} belongs to the un-correlated case between detection intensities and presence probabilities.

A justification for the above phenomena can be summarized as follows. Let $ \omega$ be the limit of $\widehat\psi_c$ as $n \rightarrow \infty$. Since $\widehat \btheta_c $ is consistent for $\btheta$, by (\ref{psi-c}), it can be shown that $\omega$ is, asymptotically, the solution of
\begin{linenomath*}
\begin{equation*} 
\frac{1}{n}\sum_{i = 1}^n \frac{\pi_i\psi_i - \omega \pi_i}{1 - \omega\pi_i} = 0. \label{eq:psi} 
\end{equation*} 
\end{linenomath*}
A first-order approximation then implies that $\left(\overline{\psi \pi} - \omega \bar \pi\right)/\left(1 - \omega \bar \pi\right) \approx 0$ where $\overline{\psi \pi}$ denotes the average of $\psi_i\pi_i$. Therefore, we have $\omega \approx \overline{\psi \pi}/\bar \pi$ which supports the finding as $1 - e^{-\lambda}$ is a monotonic increasing function of $\lambda$, noting that maximum likelihood $\widehat \psi$ shows a similar behaviour as $\widehat \psi_c$ as seen in Section~\ref{sec:sim}. Moreover, as pointed by a referee, abundance and presence probabilities are generally not independent in most occupancy studies. Thus, the average presence probability estimate may be biased if $\psi_i$ are non-homogeneous.

Although $\widehat\psi_c$ is not a consistent estimator for the average presence probability $\bar \psi$ under model (\ref{model:reg2}), we propose an alternative robust (to model misspecification) estimator
\begin{linenomath*}
\[
\tilde \psi = \frac{1}{n}\sum_{i \in A} \frac{1}{1 - e^{-\widehat \lambda_i}}.
\]
\end{linenomath*}
where $\widehat \lambda_i = \exp(\widehat \btheta_c'x_i)$ and $\widehat \btheta_c$ is the conditional likelihood estimate of (\ref{reglik2}). Recall that $A$ is the set of $\{i: y_i > 0\}$ so that $n\tilde \psi = \sum_{i = 1}^n\rI(y_i > 0)/(1 - e^{-\widehat \lambda_i})$. Specifically, $\tilde \psi$ is a Horvitz--Thompson type estimator since
\begin{linenomath*}
\[
E\left\{\frac{1}{n}\sum_{i = 1}^n \frac{\rI(y_i > 0)}{1 - e^{-\lambda_i}}\right\} = \frac{1}{n} \sum_{i = 1}^n E\left[E\left\{\frac{\rI(y_i > 0)}{1 - e^{-\lambda_i}}\middle |x_i, z_i \right\} \right] = \frac{1}{n}\sum_{i = 1}^n \psi_i = \bar \psi.
\]
\end{linenomath*}

The above Horvitz--Thompson type estimator is commonly used in capture--recapture literature, the main difference being that $i$ is indexed by an individual not a site. For spatial capture--recapture models, the proposed estimator is nearly identical for counts at traps, see \citet{borchers(2008)}.

Consequently, $\tilde \psi$ preserves consistency as long as the parameter $\btheta$ can be estimated consistently, where the latter is generally true only if the detection component model is specified correctly. Moreover, as shown in Proposition~2, the estimating function of $\widehat \btheta_c$ is optimal in this case. Hence, the estimator $\tilde \psi$ preserves some robustness to model misspecification.

To estimate the variance of $\tilde \psi$, we consider an approximation by using the delta method
\begin{linenomath*}
\[ 
{\rm var}(\tilde \psi) \approx \frac{1}{n^2}\sum_{i = 1}^n \frac{\psi_i \left\{1 - \psi_i (1 - e^{-\lambda_i})\right\}}{1 - e^{-\lambda_i}} + \frac{1}{n^2}D_\btheta' {\rm var}(\widehat \btheta)D_\btheta 
\]
\end{linenomath*}
where $D_\btheta = \sum_{i \in A}e^{-\lambda_i}\lambda_i x_i/(1 - e^{-\lambda_i})^2$ and ${\rm var}(\widehat \btheta)$ is the inverse observed Fisher information of (\ref{reglik2}). Since the $\psi_i$'s are unknown, we propose the following variance estimator
\begin{linenomath*}
\[ 
\widehat{\rm var}(\widehat \psi) = \frac{1}{n^2}\sum_{i \in A} \frac{1 - \tilde \psi (1 - e^{-\widehat{\lambda}_i})}{(1 - e^{-\widehat\lambda_i})^2} + \frac{1}{n^2} D_\btheta' {\rm var}(\widehat \btheta)D_\btheta
\]
\end{linenomath*}
where $\btheta$ is evaluated at $\widehat \btheta$.

The estimator $\tilde \psi$ is motivated by similar estimators used for species richness estimation and closed population capture--recapture models. The estimator $n \tilde \psi$ is a population size estimator \citep{van} for the number of occupied sites (with some sites having zero observed counts). However, to the best of our knowledge, a population size estimator of this type that uses an estimate of the average presence probability $\bar \psi$ has not yet been considered in the literature. Moreover, the variance estimator of $n \tilde \psi$ commonly used in the capture--recapture literature cannot be applied directly here where we have to estimate $\psi_i$ as mentioned above. 
%In Table~\ref{tab_bias}, we summarize the findings for the behaviour of the presence probability estimator. 

%\begin{table}[ht]
%\centering
%\caption{\it A summary for the behaviour of $\widehat \psi$. Here, we denote $\psi$ and $\lambda$ as constants, and, $\lambda_i$ and $\psi_i$ as functions of covariates, $\rho = \{1 + 0.5\sigma^2/(e^{\mu} - 1 - \mu)\}^{-1}$. The $c0$, $c+$ and $c-$ represent the cases of zero, positive, and negative correlations between presence probabilities and detection intensities, respectively. \label{tab_bias}}
%\begin{tabular}{cc|c|c}
%\hline
%\hline
%\multicolumn{2}{c|}{True Model} & Working Model & Behaviour of $\widehat{\psi}$\\ 
%\hline
%\multicolumn{1}{c|}{} & $\lambda \sim g_\theta$ & & \\ 
%\cline{2-2}
%\multicolumn{1}{c|}{\multirow{-2}{*}{$\psi$}} & $\lambda_i$ & \multirow{-2}{*}{$\psi, \lambda$} & \multirow{-2}{*}{$\widehat{\psi} < %\psi$} \\ \hline
%\multicolumn{2}{c|}{$\psi, {\scriptstyle {\rm E} (\lambda) = \mu, {\rm var} (\lambda) = \sigma^2}$} & $\psi, \lambda$ & $\widehat \psi %\approx \rho\psi$ \\ 
%\hline
%\multicolumn{1}{c|}{} & $c0$ & & $\widehat{\psi} \approx \bar{\psi}$ \\ 
%\cline{2-2} \cline{4-4}
%\multicolumn{1}{c|}{} & $c+$ & & $\widehat{\psi} > \bar{\psi}$ \\ 
%\cline{2-2} \cline{4-4}
%\multicolumn{1}{c|}{\multirow{-3}{*}{$\psi_i, \lambda_i$}} & $c-$ & \multirow{-3}{*}{$\psi, \lambda_i$} & $\widehat{\psi} < \bar{\psi}$ \\ 
%\hline
%\end{tabular}
%\end{table}

\section{Presence--absence data: binomial models}
\label{sec:binary}

Following similar notation as presented in Section~2, suppose now that $y_{ij} = 1 $ or $0$ indicates the presence/absence of some species at site $i = 1, \ldots, n$ and occasions (or visits) $j = 1, \ldots, T$. For site $i$, denote the total observed frequency by $y_i = \sum_{j = 1}^T y_{ij}$. Assume that $y_{i} \sim (1 - \psi)\rI(0) + \psi {\rm Bin}(T, p)$ -- \ie it has a zero-inflated binomial (ZIB) distribution with detection probability $p$ and the presence probability $\psi$. We again let $m_k = \sum_{i = 1}^n \rI(y_i = k)$ for all $k \geq 0$ and let $m_+ = \sum_{k \geq 1} m_k = n - m_0$. The likelihood function for estimating $(p, \psi)$ is
\begin{linenomath*}
\begin{equation}
L(p, \psi) \propto \prod_{k \geq 1} \left \{\psi p^k(1 - p)^{T - k}\right\}^{m_k} \times \left \{(1 - \psi) + \psi (1 - p)^T \right\}^{m_0}. \label{eq:pa1}
\end{equation}
\end{linenomath*}

We again refer to the above likelihood function as the {\it homogeneous occurrence model}. As in the previously discussed abundance models, we can extend the homogeneous occurrence model to include detection heterogeneity as 
\begin{linenomath*}
\begin{equation}
L(\btheta, \psi) = \left\{\prod_{k = 1}^T p_k(\btheta)^{m_k}\right\} \psi^{m_+} \times\{(1 - \psi) + \psi p_0(\btheta)\}^{m_0}. \label{eq:pa2}
\end{equation}
\end{linenomath*}
where $p_k(\btheta) = {T \choose k} \int_0^1 p^k(1 - p)^{T - k} g_\btheta(p) {\rm d} p$ and $g_\btheta(p)$ is a probability density function (or probability mass function) of $p$ on $(0, 1)$. For example, we could set $g_\btheta(p)$ to follow a beta distribution as in \citet{royle}. The estimation procedure for conditional likelihood can also be applied to model (\ref{eq:pa2}). In doing so, the conditional likelihood method shares the same advantage as in~\citet{morgan} who proposed a re-parametrization that simplifies the likelihood making computation/model fitting a lot easier. It turns out that the approach of \citet{morgan} can be viewed as a special case of conditional likelihood. In addition to model (\ref{eq:pa2}), the inclusion of covariates in detection probabilities and/or presence probabilities is also straightforward, \eg $p_i = H(\btheta'x_i) $ and/or $\psi_i = H(\gamma'z_i)$. This is analogous to abundance models (\ref{reglik}) and (\ref{model:reg2}).

For occurrence models, similar results to Propositions~1--\ref{prop7} can be established. We omit most of these results since they are similar to those given in the previous sections. An exception, however, is the approximate bias (Proposition~1) which we present in the Web Appendix.

Finally, if detection probabilities are $p_i = H(\btheta'x_i)$ for some covariates $x_i$, the aforementioned Horvitz--Thompson type estimator $\tilde \psi$ can be defined as $\tilde \psi = n^{-1}\sum_{i \in A} \widehat \pi_i^{-1}$ where $\widehat \pi_i = 1 - H(-\widehat \btheta_c' x_i)^T$, and $\widehat \btheta_c$ is the conditional likelihood estimate \citep{huggins89}. In this case, the corresponding variance estimator is
\begin{linenomath*}
\[ 
\widehat {\rm var}(\widehat \psi) = \frac{1}{n^2}\sum_{i \in A} \frac{1 - \tilde \psi \pi_i}{\pi_i^2} + \frac{1}{n^2} \left \{\sum_{i \in A} \frac{T p_i (1 - \pi_i) x_i}{\pi_i^2} \right \}'{\rm var}(\widehat \btheta_c) \left \{\sum_{i \in A} \frac{T p_i (1 - \pi_i)x_i}{\pi_i^2} \right\}
\]
\end{linenomath*}
with $\btheta$ is evaluated at $\widehat \btheta_c$. 

\section{Simulations}
\label{sec:sim}

We conducted various simulation studies to verify the performance of conditional likelihood and maximum likelihood estimation, and to investigate the impacts of ignoring heterogeneity in detection and presence probabilities. For each simulation scenario, we generated 1,000 data sets.

\subsection{Scenario (a)}
\label{subsec:sim1a}

First, we examined the effects of ignoring heterogeneity in detection, see Section~\ref{subsec:eff_det}. We generated data from a zero-inflated negative binomial distribution. Let NB$(\kappa, \mu/\kappa)$ denote a negative binomial distribution with a dispersion parameter $\kappa$ and mean $\mu$, such that, NB$(\kappa, \mu/\kappa)$ reduces to a Poisson($\mu$) as $\kappa$ is increased to infinity, and NB$(\kappa, \mu/\kappa)$ is equivalent to a mixture Poisson$(\lambda)$ where $\lambda \sim {\rm Gamma}(\kappa, \mu/\kappa)$. Hence, the mixing distribution has a mean $\mu$ and variance $\mu^2/\kappa$, and this allows us to obtain the approximate bias as given in Proposition~1. We investigated the estimation of $\psi$ using the homogeneous abundance zero-inflated Poisson model. In this simulation scenario, conditional likelihood estimates are exactly equal to maximum likelihood estimates (by Lemma 1), so we only presented maximum likelihood estimates for zero-inflated negative binomial models, which were also used as our reference model.

\begin{figure}
\includegraphics[height=100mm, width=155mm]{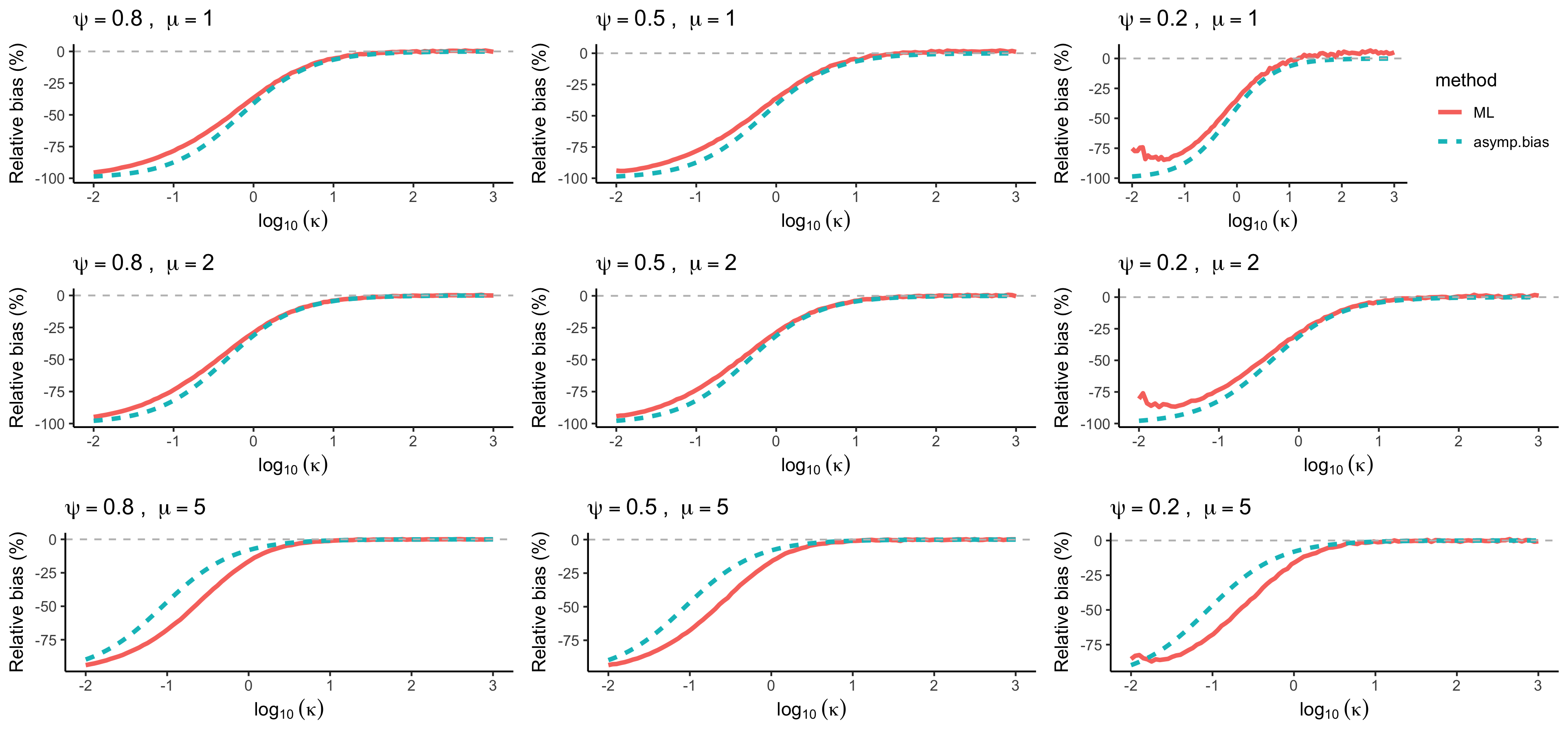}
\caption{\it Simulation Scenario (a): Asymptotic (relative \%) bias for $\psi$ (dashed green line) and bias for $\psi$ when using maximum likelihood (ML, red solid line) as a function of $\log_{10}\kappa$. We set $\mu = 1, 2, 5$, $n = 200$ and $\kappa = 0.01, \ldots, 1000$.} \label{fig1}
\end{figure}

We set $\psi = 0.2, 0.5, 0.8$, $\mu = 1, 2, 5$, $n = 200$ and $\kappa = \{ 10^{-2}, \ldots, 10^3\}$ which is an evenly spaced sequence on a log scale of length 100. In Figure~\ref{fig1} we plot the relative \% bias for $\psi$ when using the maximum likelihood estimator and compare this with asymptotic (relative \%) bias for $\psi$ as a function of $\kappa$. These results demonstrate the underestimation property when detection heterogeneity is ignored. Moreover, a general agreement is observed between the approximate bias (Proposition 1) and the empirical bias, in particular, this is noticeable when the variance $\mu/\kappa^2$ decreases.

\subsection{Scenario (b)}
\label{subsec:sim1b}

Next, we introduced covariates in the detection component (\ie regression models) and examined model performance (Section~\ref{subsec:reg}). We let $x_i = (1, x_{1 i}, x_{2 i})^T$ where: (i){ $x_{1 i}$ and $x_{2 i}$ are independent standard normal random variables}; and (ii) $x_{1 i} \sim {\rm Bern}(0.5)$ and $x_{2 i} \sim \mathcal{N}(0, 1)$. Thus, we considered $\lambda_i = \exp(\theta_0 + \theta_1 x_{1 i} + \theta_2 x_{2 i})$, and estimated $\bm \theta = (\theta_0, \theta_1, \theta_2)$ and $\psi$ with the following model structures for the detection component: a constant only, the covariate $x_{1 i}$ only, and both covariates $x_{1 i}$ and $x_{2 i}$ (correct structure). We denote these models by Model $\cdot$, $x_1$ and $x_1 + x_2$, respectively. We fitted each model with these structures using conditional likelihood and maximum likelihood.

To evaluate the performance of conditional and maximum likelihood in the regression models (Proposition \ref{prop5}), we compared estimates for $\bm \theta$ when two covariates $x_1$ and $x_2$ are included in the detection component (Model $x_1 + x_2$). We set $\bm{\theta} = (1, -1, 1)$, $\psi = 0.75$ and $n = 200$. Results for $\btheta$ are presented in Table~\ref{tab1}, and for $\psi$ in Table~\ref{tab2} when fitting Model $x_1 + x_2$. We report the average of estimates, the sample standard deviation, the average of the standard error estimates, the root mean square error, and the sample coverage percentage of the~95\% confidence intervals of estimators. In Table~\ref{tab2}, we compared estimates for $\psi$ when fitting Models $\cdot$ and $x_1$ (\ie when detection heterogeneity is completely or partially ignored).

According to the results of Table~\ref{tab1}, maximum likelihood outperforms conditional likelihood in terms of bias and root mean square error, but the differences are generally quite minor, especially for case (i). Nevertheless, we noticed that for the coverage percentage criterion, maximum likelihood estimation achieved the nominal 95\% level closer than conditional likelihood estimation. Specifically, the sample coverage percentages of conditional likelihood were only around 90--93\% for the non-intercept regression parameters $\theta_1$ and $\theta_2$. The reason for this is due to the standard error estimator of conditional likelihood yielding a negative bias for finite samples; thus, there were some gaps between the standard deviation and standard error estimates. A bootstrap confidence interval could improve the interval estimation, but this requires further investigation. Additionally, in Table~\ref{tab2}, we once again see that the effects of ignoring heterogeneity in detection will yield a negative bias for estimating the presence probability. We also found that the downward bias effect remains only if some covariates are ignored; that is, notice the average of estimates by Model $x_1$ in Table~\ref{tab2}.

\begin{table}[ht]
\centering
\begin{threeparttable}
\caption{\it Simulation Scenario (b): Estimates for $\bm \theta = (\theta_0, \theta_1, \theta_2)$ when fitting Model $x_1 + x_2$ and using maximum likelihood (ML) and conditional likelihood (CL) for cases (i)--(ii). We set $\bm{\theta} = (1, -1, 1)$ and $n = 200$}{
\begin{tabular}{llrrrrrr}
\hline
\hline
Case (i) & Parameter& \multicolumn{2}{c}{$\theta_0$} & \multicolumn{2}{c}{$\theta_1$} & \multicolumn{2}{c}{$\theta_2$} \\
& Method & ML & CL & ML & CL & ML & CL \\ 
\hline
& AVE & 0.997 & 0.998 & -1.000 & -1.000 & 1.001 & 1.001 \\ 
& SD & 0.060 & 0.065 & 0.035 & 0.036 & 0.036 & 0.038 \\ 
& A.SE & 0.059 & 0.061 & 0.035 & 0.034 & 0.035 & 0.034 \\ 
& RMSE & 0.060 & 0.065 & 0.035 & 0.036 & 0.036 & 0.038 \\ 
& CP & 94.30 & 93.30 & 95.80 & 93.00 & 93.50 & 90.60 \\ 
\hline
\hline
Case (ii) & Parameter& \multicolumn{2}{c}{$\theta_0$} & \multicolumn{2}{c}{$\theta_1$} & \multicolumn{2}{c}{$\theta_2$} \\
& Method & ML & CL & ML & CL & ML & CL \\ 
\hline
& AVE & 0.993 & 0.991 & -1.000 & -1.003 & 1.001 & 1.003 \\ 
& SD & 0.082 & 0.090 & 0.116 & 0.125 & 0.058 & 0.065 \\ 
& A.SE & 0.080 & 0.085 & 0.115 & 0.120 & 0.054 & 0.057 \\ 
& RMSE & 0.082 & 0.090 & 0.116 & 0.125 & 0.058 & 0.065 \\ 
& CP & 94.70 & 94.60 & 94.90 & 93.20 & 94.70 & 90.10 \\ 
\hline
\end{tabular}
\footnotesize
\begin{tablenotes}
\item[] AVE, Average of estimates; SD, sample standard deviation; A.SE, average of standard error estimates; RMSE; root mean square error; CP, sample coverage percentage 95\% confidence intervals
\end{tablenotes}}
\label{tab1}
\end{threeparttable}
\end{table}

\begin{table}[ht]
\centering
\begin{threeparttable}
\caption{\it Simulation Scenario (b): Estimates for $\psi$ when using maximum likelihood and conditional likelihood for cases (i) and (ii) under three detection models (Models $\cdot$, $x_1$ and $x_1\!+\!x_2$). The three models denote the constant only, using covariate $x_1$ and $x_1\!+\!x_2$ in the detection component respectively. We set $\psi = 0.75$ and $n = 200$}{
\begin{tabular}{llrrrrrrr}\\
\hline
\hline
& Method & \multicolumn{3}{c}{Maximum likelihood} & & \multicolumn{3}{c}{Conditional likelihood}\\
\hline
Case (i) & Model & $\cdot$ & $x_1$ & $x_1\!\!+\!\!x_2$ & & $\cdot$ & $x_1$ & $x_1\!\!+\!\!x_2$ \\ 
\hline
& AVE & 0.595 & 0.640 & 0.748 & & 0.595 & 0.638 & 0.747 \\ 
& SD & 0.034 & 0.041 & 0.037 & & 0.034 & 0.041 & 0.037 \\ 
& A.SE & 0.035 & 0.037 & 0.039 & & 0.035 & 0.040 & 0.039 \\ 
& RMSE & 0.158 & 0.117 & 0.037 & & 0.158 & 0.119 & 0.037 \\ 
& CP & 0.5 & 16.2 & 95.5 & & 0.5 & 19.4 & 95.4 \\ 
\hline
Case (ii) & Model & $\cdot$ & $x_1$ & $x_1\!\!+\!\!x_2$ & & $\cdot$ & $x_1$ & $x_1\!\!+\! \!x_2$ \\ 
\hline
& AVE & 0.560 & 0.587 & 0.753 & & 0.560 & 0.584 & 0.752 \\ 
& SD & 0.035 & 0.039 & 0.042 & & 0.035 & 0.039 & 0.042 \\ 
& A.SE & 0.036 & 0.038 & 0.044 & & 0.036 & 0.040 & 0.045 \\ 
& RMSE & 0.193 & 0.168 & 0.042 & & 0.193 & 0.171 & 0.042 \\ 
& CP & 0 & 1.8 & 95.5 & & 0 & 1.3 & 96.2 \\ 
\hline
\end{tabular}
\footnotesize
\begin{tablenotes}
\item[] AVE, Average of estimates; SD, sample standard deviation; A.SE, average of standard error estimates; RMSE; root mean square error; CP, sample coverage percentage 95\% confidence intervals
\end{tablenotes}}
\label{tab2}
\end{threeparttable}
\end{table}

\subsection{Scenario (c)}
\label{subsec:sim1c}

Lastly, we investigated the effects of ignoring heterogeneity in the presence probabilities. The main interest here is in examining the average presence probability $\bar \psi$. We used a similar set up as Scenario (b) but now considered both detection and presence probabilities as regression models. Specifically, we let $\lambda_i = \exp(\theta_0 + \theta_1 x_{i1} + \theta_2 x_{2 i})$ where $x_{1 i} \sim N(0, 1)$ and $x_{2 i} \sim Ber(0.5)$, and let $\psi_i = 1/\{1 + \exp(-\gamma_0 - \gamma_1 x_{1 i}\}$. We generated the data by fixing $\bgamma = (1, 1)$ and three cases of $\btheta$, namely $\btheta = (1, 1, 1), (1, 0, 1), (1, -1, 1)$. We label the cases as $c+$, $c0$ and $c-$, see Table~1, as they represent positive, zero, and negative correlations between presence probabilities and detection intensities. 

As in Scenario (b), we considered different occupancy models but now assumed the detection structure was correctly specified. We fitted the following models where the occurrence component was modelled as: constant only, and the covariate $x_{1}$. These are denoted as Model $\cdot$ and Model $x_1$ in Table~\ref{tab3}. We fitted each of these model structures using maximum likelihood, conditional likelihood and calculated the proposed Horivtz--Thompson estimator $\tilde \psi$ (denoted by conditional likelihood$^*$).

Once again, we observed similar performances for $\btheta$ for maximum likelihood and conditional likelihood, so the results are not reported here. In Table~\ref{tab3} we give the results for $\bar \psi$. From the reduced model, we see that $\widehat\psi$ and $\widehat\psi_c$ showed a positive, negligible, and negative bias for cases $c+$, $c0$, and $c-$, respectively. However, the estimator showed some consistency when the occurrence component was correctly specified (\ie Model 2). Finally, we see that the Horvitz--Thompson estimator $\tilde \psi$ serves as a satisfactory robust method, as the proposed standard error estimator also performed well with small positive biases.

\begin{table}[ht]
\centering
\begin{threeparttable}
\caption{\it Simulation Scenario (c): Estimates for $\bar{\psi} = 0.681$ when using maximum likelihood (ML) and conditional likelihood (CL) for $c+$ (top), case $c0$ (middle), and case $c-$ (bottom) under two occurrence component models (Models $\cdot$ and $x_1$). Here, the detection component model is correctly specified. Method CL$^*$ uses the conditional likelihood with the Horvitz--Thompson estimator $\tilde \psi$, where the results are not affected by the specification of occurrence component model}{
\begin{tabular}{lclrrrrrrrrrr}
\hline
\hline
& & Method & \multicolumn{2}{c}{ML} & & \multicolumn{2}{c}{CL} & & \multicolumn{1}{c}{CL$^*$} & & & \\
\hline
Case $c+$ & & Model & $\cdot$ & $\cdot$ & & $\cdot$ & $x_1$ & & $\cdot$ & & &\\ 
& & AVE & 0.743 & 0.683 & & 0.737 & 0.683 & & 0.683 & & &\\ 
& & SD & 0.036 & 0.035 & & 0.036 & 0.035 & & 0.037 & & &\\ 
& & A.SE & 0.038 & 0.035 & & 0.038 & 0.035 & & 0.039 & & &\\ 
& & RMSE & 0.072 & 0.035 & & 0.067 & 0.035 & & 0.037 & & &\\ 
& & CP & 60.40 & 95.60 & & 67.60 & 95.40 & & 96.50 & & &\\ 
\hline
Case $c0$ & & Model & $\cdot$ & $x_1$ & & $\cdot$ & $x_1$ & & $\cdot$ & & &\\ 
& & AVE & 0.682 & 0.683 & & 0.682 & 0.683 & & 0.683 & & &\\ 
& & SD & 0.033 & 0.033 & & 0.033 & 0.033 & & 0.033 & & &\\ 
& & A.SE & 0.035 & 0.032 & & 0.035 & 0.032 & & 0.035 & & &\\ 
& & RMSE & 0.033 & 0.033 & & 0.033 & 0.033 & & 0.033 & & &\\ 
& & CP & 96.60 & 95.30 & & 96.90 & 95.30 & & 96.80 & & &\\ 
\hline
Case $c-$ & & Model & $\cdot$ & $x_1$ & & $\cdot$ & $x_1$ & & $\cdot$ & & &\\ 
& & AVE & 0.646 & 0.682 & & 0.648 & 0.681 & & 0.682 & & &\\ 
& & SD & 0.032 & 0.032 & & 0.032 & 0.032 & & 0.040 & & &\\ 
& & A.SE & 0.036 & 0.034 & & 0.036 & 0.034 & & 0.043 & & &\\ 
& & RMSE & 0.047 & 0.032 & & 0.046 & 0.032 & & 0.040 & & &\\ 
& & CP & 87.30 & 95.10 & & 88.40 & 95.30 & & 96.60 & & &\\ 
\hline
\end{tabular}
\footnotesize
\begin{tablenotes}
\item[] AVE, Average of estimates; SD, sample standard deviation; A.SE, average of standard error estimates; RMSE; root mean square error; CP, sample coverage percentage 95\% confidence intervals
\end{tablenotes}}
\label{tab3}
\end{threeparttable}
\end{table}

In the Web Appendix we present results for the same simulation scenarios as above but when fitting binomial models (Section~\ref{sec:binary}) to presence--absence data. Overall, we found that the results for binomial models were similar to the results given in the simulation study above.

\section{Applications}
\label{sec:ex}

We examined the effects of ignoring heterogeneity in the presence and detection probability when fitting site occupancy models to real data. In the first example, we consider a traditional site occupancy modelling approach where the data used consists of presence--absences of fish. For our second example, we used traffic violations data collected on motorcyclists in Taiwan and modelled these using a zero-inflated Poisson model as they commonly applied for the analysis of traffic accident data \citep{shankar}. The aim in both analyses is to fit the methods developed in Section~\ref{sec:model}, and to examine their performances when ignoring heterogeneity and estimating the average presence probability.

\subsection{Brook trout data}
\label{subsec:ex1}

These data were collected by Professor James T. Peterson via electro-fishing, and were analysed in \citet{huggins19}. Sampling was conducted on 50m sections of streams at 57 sites in the Upper Chattahoochee 371 River basin, USA. The number of sites was $77$ with $3$ sampling occasions. The frequencies for counts of $Y = 0, \ldots, 3$ were 45, 11, 17, and 4, so the observed presence percentage was 32/77 = 41.6\%. Two covariates were identified to be significantly associated with detection and presence probabilities \citep[see][]{huggins19}: the elevation ($elev$), which was measured in kilometres, and the stream mean cross-sectional area ($CSA$). We took the average of $CSA$ measurements over three visits for each site.

We used $CSA$ and elevation to model the detection and occurrence components, respectively. We present the estimates of each coefficient when using maximum likelihood and conditional likelihood in Table~\ref{tab4}. In this example, maximum likelihood and conditional likelihood showed some differences in the detection component, $logit(p)$, where the effect of $CSA$ is of lesser significance for conditional likelihood. However, results were more consistent for the occurrence component, $logit(\psi)$.
 
Next, we postulate that the above (full) model is true, and we removed the covariates (\ie the modelled heterogeneity) from the occurrence and detection components to see the impacts of estimating the {average presence probability} $\bar\psi$. We report the results in Table~\ref{tab4}. First, the average presence probability is estimated at 48.6\% by maximum likelihood; hence, it implies that about 7\% of occupied sites were not detected. By removing the heterogeneity in the detection component (see model Det.\ const.), maximum likelihood estimate of $\bar\psi$ reduces by 2.6\% (which contributes to a significant portion in 7\%). Similar findings were seen for conditional likelihood and conditional likelihood$^*$, though these were less significant due to the effects of $CSA$ being of lesser significance by conditional likelihood (the degree of detection heterogeneity is more negligible in conditional likelihood). Finally, by removing the heterogeneity in the occurrence component (see model Pres.\ const.), all estimates of the three methods showed almost no change compared to the full model. The reason for this is that detection probabilities and presence probabilities are nearly uncorrelated as the correlation coefficient of $CSA$ and $elev$ is only 0.036 ($p$-value 0.75). 

\begin{table}[ht]
\centering
\begin{threeparttable}
\caption{\it (a) Model estimates and standard errors using the Brook trout data (Trout) and Taiwan motorcycle data (Motor.) by maximum likelihood (ML) and conditional likelihood (CL). Values in parentheses represent the standard error estimates.(b) Average presence probability ($\bar \psi$) estimates and standard errors using the Brook trout data and Taiwan motorcycle data when fitting conditional likelihood and maximum likelihood. Method CL$^*$ uses the conditional likelihood with the Horvitz--Thompson estimator $\tilde \psi$}{
\small
$\hspace*{-1cm}$\begin{tabular}{lrrrrrrrrrrr}
\hline
\hline
\multicolumn{1}{l}{(a)} & \multicolumn{1}{c}{Det./Pres.} & \multicolumn{5}{c}{Maximum likelihood} & \multicolumn{5}{c}{Conditional likelihood}\\
\hline
\multicolumn{1}{l}{Trout} & \multicolumn{1}{c}{$logit(p)$} & \multicolumn{5}{c}{$ \tb{1.458}{(0.687)} - \tb{0.914}{(0.431)}\!\!$ {\footnotesize $CSA$}} & \multicolumn{5}{c}{$\tb{1.157}{(0.770)} - \tb{0.694}{(0.465)}\!\!$ {\footnotesize $CSA$}}\\ 
& \multicolumn{1}{c}{$logit(\psi)$} & \multicolumn{5}{c}{$\tb{-4.501}{(1.185)} + \tb{1.526}{(0.404)} \!\!$ {\footnotesize $elev$}} & \multicolumn{5}{c}{$\tb{-4.438}{(1.129)} + \tb{1.489}{(0.357)} \!\!$ {\footnotesize $elev$}} \\
\hline
\multicolumn{1}{l}{Motor.} & \multicolumn{1}{c}{$log(\lambda)$} & \multicolumn{5}{c}{$ \!\tb{-0.172}{(0.084)} \! + \!\!\!\tb{0.284}{(0.123)}\! \!\!$ {\footnotesize $km$} $\!+\!\! \!\tb{0.193}{(0.109)} \!\!\!$ {\footnotesize $eng$}} & \multicolumn{5}{c}{$\! \tb{-0.187}{(0.089)}\! +\!\!\! \tb{0.326}{(0.129)}\!\! \!$ {\footnotesize $km$} $\! \!+\!\!\! \tb{0.205}{(0.109)}\! \! \!$ {\footnotesize $eng$}} \\ 
& \multicolumn{1}{c}{$logit(\psi)$} & \multicolumn{5}{c}{$\!\tb{-2.074}{(0.092)} \!+ \!\!\!\tb{0.510}{(0.163)}\! \!\!$ {\footnotesize $km$} $\!+\!\! \!\tb{2.055}{(0.164)} \!\!\!$ {\footnotesize $eng$}} & \multicolumn{5}{c}{$ \tb{-2.063}{(0.096)}\! +\!\!\! \tb{0.476}{(0.166)}\!\! \!$ {\footnotesize $km$} $\! \! +\!\!\! \tb{2.043}{(0.162)}\! \! \!$ {\footnotesize $eng$}} \\ 
\hline
\hline
\multicolumn{1}{l}{(b)} & \multicolumn{3}{c}{Full model} & & \multicolumn{3}{c}{Pres. const. model} & & \multicolumn{3}{c}{Det. const. model} \\
\hline
& ML & CL & CL$^*$ & & ML & CL & CL$^*$ & & ML & CL & CL$^*$ \\ 
\hline
\multicolumn{1}{l}{Trout} & $\tb{0.486}{(0.057)}$ & $\tb{0.479}{(0.058)}$ & $\tb{0.472}{(0.070)}$ & & $\tb{0.480}{(0.071)}$ & $\tb{0.476}{(0.068)}$ & $\tb{0.472}{(0.070)}$ & & $\tb{0.460}{(0.055)}$ & $\tb{0.461}{(0.056)}$ & $\tb{0.463}{(0.066)}$\\
\hline
\multicolumn{1}{l}{Motor.} & $\tb{0.175}{(0.009)}$ & $\tb{0.175}{(0.009)}$ & $\tb{0.175}{(0.009)}$ & & $\tb{0.280}{(0.018)}$ & $\tb{0.183}{(0.011)}$ & $\tb{0.175}{(0.009)}$ & & $\tb{0.174}{(0.008)}$ & $\tb{0.173}{(0.009)}$ & $\tb{0.173}{(0.009)}$\\
\hline
\end{tabular}
\footnotesize
\begin{tablenotes}
\item[] $CSA$, stream mean cross-sectional area; $elev$, elevation.
\item[] $km$, $\rI({\rm riding~distance} > 10,000{\rm km})$; $eng$, $\rI({\rm engine~volume} > 250 {\rm cc})$.
\item[] Full, full model; Pres.\ const., presence probability is constant; Det.\ const., detection is constant.
\end{tablenotes}}
\label{tab4}
\end{threeparttable}
\end{table}

\subsection{Motorcycle traffic violation data}
\label{subsec:ex2}

This survey study consists of 7,386 respondents who provided information on their motorcycle usage, such as, maintenance cost, the purpose of use, mileage, engine volume, and information on the owners (\ie the riders), including age and the number of traffic violations in one year. The Ministry of Transportation and Communication conducted the survey in Taiwan in 2007.

We considered traffic violations frequencies by riders as the response variable, $Y$. Due to some riders violating the traffic rules but not being detected, the zero counts of $Y$ are a mixture of random and structural zeros~\citep{yee(2015a)}. \citet{smlee} fitted a zero-inflated Poisson model with respect to two regressors: riding distance per year and motorcycle engine volume. Similarly, we define $km = \rI({\rm riding~distance} > 10,000{\rm km})$ and $eng = \rI({\rm engine~volume} > 250 {\rm cc})$. Engines exceeding 250cc are classified as heavy-motorcycles, which require a special driver’s license in Taiwan. These data contained some missing values, but their absence only affected the results slightly \citep{smlee}. In our analysis, we used the complete data set of 5,447 observations. The observed presence percentage was 10.8\%, $km = 1 (849)$ and $eng = 1 (754)$. We present model estimates when using maximum likelihood and conditional likelihood in Table~\ref{tab4}. For this example, maximum likelihood and conditional likelihood showed similar estimates in both the $log(\lambda)$ and $logit(\psi)$ components.

As in Section~\ref{subsec:ex1}, we postulate that the above (full) model is true, and we removed the model heterogeneity from the presence probability and detection component. We report the results in Table~\ref{tab4}. First, the average presence probability is estimated at 17.5\% by maximum likelihood; this implies that an additional 6.7\% of the traffic violation drivers were not detected. Next, by removing the heterogeneity in the detection component (see model Det.\ const.), all three estimates of $\bar\psi$ were reduced by 0.2\%. This may be due to low presence percentages in the data. However, by removing the heterogeneity in the occupancy component (see model Pres.\ const.), the maximum likelihood estimate is increased to 28\% (with an increment of 10.5\% \emph{c.f.}\ the full model). The reason for this is detection and presence probabilities are highly positively correlated, where both components use the same covariates, and all have significant positive effects. Here, the average presence probability estimate for conditional likelihood had also increased, but this was not as profound. Again, we see that the average presence probability estimate for conditional likelihood$^*$ is not affected by the specification of the occurrence component.

\section{Discussion}
\label{sec:dis}

In this study, we examined the performance of parameter estimates in occupancy models when there is heterogeneity in presence and detection probabilities. We presented three key findings: (1.) the conditional maximum likelihood is not asymptotically equivalent to the usual maximum likelihood estimator for regression models, (2.) the presence probability is underestimated if detection heterogeneity is unaccounted for, and (3.) we examined the effects of estimating the average species presence probability if presence heterogeneity is unaccounted for. The results for (1.) and (2.) were established for constant presence probabilities. However, similar results may be extended to non-homogeneous presence probabilities. For (3.), we have only given partial results when detection heterogeneity is considered to be a mixture model or when the conditional likelihood approach is used in a regression framework. Results for the maximum likelihood estimator in the regression setting still require further research.
 
Although we showed that conditional likelihood and maximum likelihood are not asymptotically equivalent in the regression case, most of our empirical results showed similarities in terms of efficiency. In practice, maximum likelihood is widely used; however, conditional likelihood has a much simplified form which makes computation/model fitting a lot easier, and if covariates are used, it benefits from having an optimal estimating function property which ensures robustness against model misspecification on the presence probability (Proposition 2). Hence, we also recommend the use of conditional likelihood in occupancy studies.

We proposed a Horvitz--Thompson type estimator for the average presence probability based on conditional maximum likelihood. This is a robust method only if the detection component model is correctly specified. However, there is no general method to check for model misspecification. In practice, these estimates can serve as a tool in providing references estimates for checking whether an occupancy model is suitable. Occasionally, we found that the Horvitz--Thompson estimator and the associated confidence intervals may yield unreasonable values, such as estimates greater than one, and the lower bound of the confidence interval is estimated to be smaller than the observed presence probability ($m_+/n$). \citet{eren} proposed a modified log-transformation confidence interval that deals with the doubly-bounded situation when the point estimator is smaller than 1. However, there is currently no literature on implementing a constrained Horvitz--Thompson estimator with values being less than one.

This study specifically focused on the effects of ignoring heterogeneity among sites in the occupancy models. Indeed, heterogeneity is often the most critical factor that affects the model in practice. However, additional factors, such as temporal variation and dependence among visits, should also be considered in the model. Examining these additional effects have been similarly considered in capture--recapture models, see \citet{hwang05, rivest}. We envisage that similar results could be developed in the occupancy modelling context provided that the presence probability is constant; however, this can be very challenging when heterogeneity in the presence probabilities is also included. In addition, there may be interactions between these factors, which occurs in practice, and therefore adding further complexity to the problem.

\section*{Acknowledgement}
\label{sec:ack}

%The authors would like to thank the associate editor and a referee for their valuable comments. 
We would like to thank Prof.\ James T.\ Peterson and Prof.\ Shen-Ming Lee for providing data sets used in Section \ref{sec:ex}. We would also like to thank Dr.\ Gurutzeta Guillera-Arroita for providing comments to an earlier version of this manuscript. This work was supported by the Ministry of Science and Technology of Taiwan.

\end{document}